\documentclass[%
reprint,
nofootinbib,
 amsmath,amssymb,
 aps,
prl,
]{revtex4-2}

\usepackage{graphicx}
\usepackage{dcolumn}
\usepackage{bm}
\usepackage{color}

\usepackage{subfigure}

\begin{document}

\preprint{APS/123-QED}

\title{Gravitational-wave Tomography of the Moon:\\ Constraining Lunar Structure with Calibrated Gravitational Waves}

\author{Han Yan$^{1,2,3}$}
    \email{Corresponding author. \\ yanhanphy@pku.edu.cn, hyan.phy@gmail.com}
\author{Jan Harms$^{3,4}$}

\affiliation{$^{1}$Department of Astronomy, School of Physics, Peking University, 100871 Beijing, China}

\affiliation{$^{2}$Kavli Institute for Astronomy and Astrophysics, Peking University, 100871 Beijing, China}

 \affiliation{$^{3}$Gran Sasso Science Institute, 67100 L'Aquila, Italy}

\affiliation{$^{4}$INFN, Laboratori Nazionali del Gran Sasso, 67100 Assergi, Italy}

\date{\today}

\begin{abstract}
The recent success of gravitational-wave (GW) astronomy together with renewed plans for lunar geophysical instrumentation has revived interest in using the Moon as a resonant detector for mid-frequency (mHz-Hz) GWs.  In realistic observational scenarios, the GW strain amplitude is expected to be constrained independently by networks of GW detectors, which motivates an inverse, \emph{tomographic} question: to what extent can measurements of the Moon's seismic response to known GWs be used to infer its internal structure? In this work, we develop a first-principles, perturbative framework that maps spherically symmetric perturbations of the elastic and density structure to measurable changes in observables, especially GW-driven modal amplitudes of the Moon. The formalism combines (i) a normal-mode representation of the elastic response, (ii) first-order perturbation theory for eigenvalues and eigenfunctions, and (iii) a linearized observation model that links frequency and amplitude observables to model parameters (bulk and shear moduli, density, and interface locations) and their perturbations. We show that the estimation errors of the Moon's elastic parameters can be reduced by about an order of magnitude with observations of calibrated GWs. 
\end{abstract}

\keywords{Moon, normal modes, gravitational waves, inverse problem, Fisher matrix, LGWA}

\maketitle


\textit{Introduction.}---Monitoring normal-mode excitations of a massive elastic body caused by gravitational waves (GWs) is an idea dating back to Weber's early proposals for a GW antenna \cite{1960PhRv..117..306W}.  This concept reemerged in recent years due to the renewed interest in the scientific utilization of the Moon.  On the one hand, renewed lunar exploration initiatives and proposals for geophysical instrumentation on the Moon have greatly improved the feasibility of long-duration, low-noise seismic monitoring \cite{10.1093/nsr/nwad329,2024LPICo3040.2354P}.  On the other hand, the concept of a future GW observatory on the Moon, including the Lunar Gravitational-Wave Antenna (LGWA) \cite{2021ApJ...910....1H,2025JCAP...01..108A}, Laser Interferometer Lunar Antenna (LILA) \cite{2024LPICo3063.5043T,2025arXiv250915452P} and related instrument studies \cite{2023JAP...133x4501V,2023SCPMA..6609513L}, envisions a lunar observatory that exploits the Moon's exceptionally quiet seismic environment to probe the mHz--Hz GW band \cite{PhysRevLett.129.071102}, complementary to both ground-based and space-based interferometers \cite{2020JCAP...03..050M,2024arXiv240207571C, 2016CQGra..33c5010L, 2021PTEP.2021eA108L}. These developments motivate revisiting the Moon as a resonant GW detector and quantifying the scientific returns possible from dedicated lunar seismometer networks.

The lunar response to GWs has been approached with a variety of methods. Theoretical efforts have revisited the lunar response using plane-wave models \cite{1969ApJ...156..529D,PhysRevD.110.064025}, Green-tensor formalism with normal-mode summation \cite{1983NCimC...6...49B,2019PhRvD.100d4048M}, and have clarified the differences and relationships among various theories \cite{PhysRevD.109.064092,2024JCAP...07..028B,2025PhRvD.111d4061M}. Complementary numerical efforts such as spectral-element models (SEMs) have started to explore detailed wave excitation and propagation in realistic lunar models \cite{2025PhRvD.111f3014Z,2026PhRvD.113b3031Z}. In parallel, instrument studies for LGWA and mission-level feasibility assessments have been developed to investigate the detectability of GW-induced signals \cite{2025JCAP...01..108A}. Collectively, these works provide the foundation to connect GW strains incident on the Moon to measurable seismic observables (modal amplitudes and eigenfrequency shifts) under increasingly realistic elastic models.

These forward-modeling advances naturally raise an inverse question: what aspects of the Moon's internal structure can be inferred from the GW-induced seismic response? In terrestrial geophysics, two complementary paradigms have been developed to tackle internal-structure inference. At long periods, the normal-mode perspective exploits high-precision frequency measurements to constrain large-scale structure \cite{1984JGR....89.5953W}, while at shorter periods, a propagating-wave perspective and full-waveform inversions target higher-resolution, three-dimensional heterogeneity \cite{1984Geop...49.1259T,2005GeoJI.160..195T}.  Historically, normal-mode-based inversions primarily used eigenfrequencies and their splitting signatures to probe internal structure (and more recently, asphericity and lateral heterogeneity) \cite{1980GeoJ...61..261W,2011GeoRL..38.4303D,2025arXiv250915452P}.

For LGWA-like observational scenarios, however, an important practical difference arises: the GW source amplitude is likely to be constrained independently and with high precision by contemporaneous GW observatories \cite{2016PhRvL.116w1102S,2020PhRvD.102b2007G,2025arXiv250710165D}, e.g., networks combining the Einstein Telescope \cite{2020JCAP...03..050M}, LISA \cite{2024arXiv240207571C}, and LGWA.  This multi-detector capability turns mode \emph{amplitude} observations, traditionally considered of limited value when source strength is \emph{unknown}, into a potentially powerful, independent probe of internal structure.  In other words, knowledge of the incident GW amplitude converts modal amplitudes into structural observables that complement frequency measurements, and opens a new inverse problem: \emph{GW tomography} of a planetary body.

We develop a largely analytical, linearized framework designed to enable forecasts and to guide numerical verification for GW-driven lunar tomography.  The formalism combines (i) a normal-mode representation of the elastic response, (ii) first-order perturbation theory for eigenfrequencies and eigenfunctions under spherically symmetric parameter perturbations, and (iii) a linearized observation model that maps frequency and amplitude observables to model parameters (bulk and shear moduli, density, and interface locations) and their perturbations. From these elements, we construct explicit Jacobian kernels and embed them in a Fisher-matrix pipeline that quantifies parameter-estimation prospects under realistic noise and prior assumptions.


\textit{Theoretical Framework.}---In this study, we model the Moon as a spherically symmetric elastic body. Self-gravity is omitted throughout this research, but can be added back in principle. The structural model is defined by radial profiles of density $\rho(r)$, bulk modulus $\kappa(r)$, and shear modulus $\mu(r)$, as well as the radii of internal discontinuities $d_i$. The free oscillations are governed by the elastic eigenvalue problem:
\begin{equation}
\label{eq:gevp}
  V\,\mathbf{s}^\sigma_{nlm} = \omega_{k}^{2}\,T\,\mathbf{s}^\sigma_{nlm},
\end{equation}
where $\sigma \in \{S,T\}$ denotes spheroidal/toroidal branches, and $\mathbf{s}^\sigma_{nlm}(\mathbf{x})$ denotes the displacement eigenfunction for a mode $k=(\sigma,n,l)$. $\omega_k$ is the eigenfrequency with $(2l+1)$-fold degeneracy. $V$ and $T$ are potential and kinetic operators respectively \cite{1999tgs..book.....D}, and are determined by the baseline one-dimensional structure of the Moon. $\omega_k$ and $\mathbf{s}^\sigma_{nlm}$ are hence also determined, and are usually calculated by a standard eigensolver such as MINEOS \cite{mineos2011}.

Under the excitation of a continuous gravitational wave (GW) with strain amplitude $h_0$ and frequency $\omega_g$, the Moon acts as a resonant detector. Previous works show that under spherical symmetry, spheroidal modes with $l=2$ are most strongly excited by GWs \cite{1983NCimC...6...49B,1996CQGra..13.2865B}. Based on the standard Green-function approach, the steady-state surface displacement for a specific mode $n$ (shorthand for $S n 2$) and for all modes (i.e., total surface response) are respectively given by \cite{1983NCimC...6...49B,2019PhRvD.100d4048M,PhysRevD.109.064092}:
\begin{eqnarray}
\label{eq:mode_amp}
\vec{\xi}_n(\mathbf{x},\omega_g) =&& h_0 \sum_{m=-2}^2 \frac{I_n f^m}{\omega_n^2-\omega_g^2 + i \omega_n \omega_g/Q_n} \mathbf{s}_{n2m}^{S}(\mathbf{x}) \nonumber \\
\vec{\xi}(\mathbf{x},\omega_g) =&& \sum_n \vec{\xi}_n(\mathbf{x},\omega_g)~,
\end{eqnarray}

where $Q_n$ is the quality factor, and $f^m$ is an $\mathcal{O}(1)$ constant pattern function depending only on the GW wave vector and polarization \cite{2019PhRvD.100d4048M}. The coupling strength is determined by the overlap integral $I_n$, which depends explicitly on the radial ($U_n$) and horizontal ($V_n$) scalar components of the eigenfunction and the shear modulus profile $\mu(r)$ (see the Supplemental Material \cite{supplemental}). At the resonant peak $\omega_g = \omega_n$, where the surface response is the strongest, the absolute amplitude is approximately $h_0 I_n U_n(R) Q_n/\omega_n^2$ for the radial response and $h_0 I_n V_n(R) Q_n/\omega_n^2$ for the horizontal response.

To quantify the constraints on the lunar interior parameters $p \in \{\rho, \kappa, \mu, d\}$, we next define a set of observables $\mathbf{O}$ indexed by the mode number $n$ and the observation type $\alpha \in \{1, 2, 3\}$:
\begin{itemize}
    \item $\alpha=1$: Eigenfrequency observables $O_{1,n}= \omega_n^2$;
    \item $\alpha=2$: Radial amplitude observables at the resonance peak $\omega_g = \omega_n$: $O_{2,n} = h_0 U_n(R)I_n \equiv A_n^r \simeq \xi_n^r \times \frac{\omega_n^2}{Q_n}$;
    \item $\alpha=3$: Horizontal amplitude observables at the resonance peak: $O_{3,n} = h_0 V_n(R)I_n \equiv A_n^h \simeq \xi_n^h \times \frac{\omega_n^2}{Q_n}$.
\end{itemize}
$U_n(R)$ and $V_n(R)$ in the definition correspond to the radial and horizontal components of $\mathbf{s}_{n2m}^{S}(\mathbf{x})$ (Eq.~\ref{eq:vecSpher}). We remove the $\frac{Q_n}{\omega_n^2}$ structure in the amplitude response based on the following considerations: first, we assume that the quality factors $Q_n$ are constrained independently (e.g., through spectral broadening measurements \cite{1978GeoJ...53..559S,1980GeoJ...61..261W,10.1111/j.1365-246X.1991.tb05700.x}) and treat them as fixed parameters; second, the measurement uncertainty of eigenfrequency $\omega_n$ is assumed to be significantly smaller than the uncertainty of the displacement measurement, i.e., $\sigma_{\{2,3\},n}/O_{\{2,3\},n} = \sigma_{A}/A_n \simeq \sigma_{\xi}/\xi_n \gg \sigma_{\omega}/\omega_n$. This is easily satisfied, as we will discuss later (before Eq.~\ref{eq:FisherInfoM}). Besides, the relative error in $h_0$ is also negligible in this work (i.e., $\sigma_{\xi}/\xi_n \gg \sigma_{h}/h_0$), because for multi-band calibration the signal-to-noise ratio (SNR) should be much better than the SNR of the lunar GW detector itself \cite{2020PhRvD.102b2007G,2025arXiv250710165D}. We also omit the $\mathcal{O}(1)$ constant factor $f^m$ and the spherical harmonics ($\mathbf{P}_{l m}$ and $\mathbf{B}_{l m}$ in Eq.~\ref{eq:vecSpher}) in ${s}_{n2m}^{S}$.

We then employ a perturbation formalism to evaluate the effect of structure variation. The sensitivity of the above observables to parameter perturbations $\delta p(r)$ ($p=\rho, \kappa, \mu, d_i$) is derived via first-order perturbation theory. Omitting the near-degeneracy effects (which can be incorporated via quasi-degenerate perturbation theory), the shift in squared eigenfrequency is given by \cite{1964ApJ...139..664C,1999tgs..book.....D}:
\begin{equation}\label{eq:delta_omega_sq}
  \delta\omega_k^2
  = \langle \mathbf{s}_{k m} \mid \delta V - \omega_k^2 \delta T \mid \mathbf{s}_{k m} \rangle ~.
\end{equation}
The inner product used above (and also in Eq.~\ref{eq:delta_s_proj}) follows the standard geophysics definition in \cite{1999tgs..book.....D}, and can be calculated through Woodhouse perturbation kernels (see Eq.~\ref{eq:WoodhouseKernel} and related discussions). Crucially, the perturbation also modifies the eigenfunctions, which in turn alters the modes $U_n, V_n$ and the overlap integral $I_n$. The first-order correction to the eigenfunction projected onto an unperturbed mode $k'\neq k$ is:
\begin{equation}\label{eq:delta_s_proj}
  \langle \mathbf{s}_{k' m'} | \delta\mathbf{s}_{k m}\rangle
  = \frac{\langle\mathbf{s}_{k' m'}|\delta V - \omega_k^2 \delta T|\mathbf{s}_{k m}\rangle}{\omega_k^2-\omega_{k'}^2}.  
\end{equation}
As a result, the final eigenfunction perturbation can be calculated as:
\begin{eqnarray}\label{eq:deltaSUV}
    \delta\mathbf{s}_{k m} &&= \sum_{k' m'} \frac{\langle\mathbf{s}_{k' m'}|\delta V - \omega_k^2 \delta T|\mathbf{s}_{k m}\rangle}{\omega_k^2-\omega_{k'}^2} \mathbf{s}_{k' m'} ~,\nonumber \\
    \delta U_{k} &&= \sum_{k' m'} \frac{\langle\mathbf{s}_{k' m'}|\delta V - \omega_k^2 \delta T|\mathbf{s}_{k m}\rangle}{\omega_k^2-\omega_{k'}^2} U_{k' } ~,\nonumber \\
    \delta V_{k} &&= \sum_{k' m'} \frac{\langle\mathbf{s}_{k' m'}|\delta V - \omega_k^2 \delta T|\mathbf{s}_{k m}\rangle}{\omega_k^2-\omega_{k'}^2} V_{k'} ~.
\end{eqnarray}
Using Eq.~(\ref{eq:deltaSUV}), we can analytically derive the Jacobian matrix elements $G_{(\alpha,n), p} = \partial O_{\alpha,n} / \partial p$ (see the details in the Supplemental Material \cite{supplemental}). 

We assume a Gaussian noise model with uncorrelated uncertainties $\sigma_{\alpha,n}$. For nominal estimates, we parameterize the relative errors as $\sigma_{1,n}/\omega_n \sim \varepsilon_\omega$ and $\sigma_{\xi}/\xi_n \sim \varepsilon_{\xi}$. While potentially optimistic for observations on the Moon, we adopt $\varepsilon_\omega\in[10^{-5},10^{-3}]$ based on long-duration terrestrial seismic benchmarks \cite{2005Sci...308.1139P}. For the relative error in the displacement measurement, we estimate $\varepsilon_{\xi}\in[10^{-2},10^{-1}]$, consistent with the proposed highly sensitive lunar GW detector \cite{2025JCAP...01..108A} where SNRs $\sim$ 100 are achievable. Therefore, the condition of $\sigma_{\{2,3\},n}/O_{\{2,3\},n} \equiv \varepsilon_A \simeq \varepsilon_{\xi} \gg \varepsilon_{\omega}$ is well satisfied \footnote{If $\varepsilon_{\omega}$ is much larger than expected, our calculations in this work need to be slightly modified. However, the importance of amplitude measurement will be even greater.}. Based on the standard procedure for a linear model, we construct the Fisher information matrix \cite{2008PhRvD..77d2001V} summing over all modes and observation types:
\begin{equation}
\label{eq:FisherInfoM}
    \mathcal{F}_{ij} = \sum_{n} \sum_{\alpha=1}^{3} \frac{1}{\sigma_{\alpha,n}^2} \frac{\partial O_{\alpha,n}}{\partial p_i} \frac{\partial O_{\alpha,n}}{\partial p_j} + (\mathbf{C}_{p}^{-1})_{ij},
\end{equation}
with a covariance prior $\mathbf{C}_{p}$ for the model parameters. We assume flat priors for the model parameters, i.e., vanishing prior information matrix, $\mathbf{C}_{p}^{-1} \to 0$.
The posterior covariance matrix is approximated by the inverse of the Fisher matrix, $\mathbf{C}_{\mathrm{post}} \approx \mathcal{F}^{-1}$. Consequently, the expected constraint on each model parameter $p_i$ is quantified by its marginal posterior standard deviation, $\sigma_{p_i} = \sqrt{(\mathbf{C}_{\mathrm{post}})_{ii}}$. 

\textit{Verification of perturbation theory.}---In the following section, we present numerical evaluations of our analytical perturbation framework on a reference lunar model, to demonstrate the resolving power of GW-induced seismic signals, especially the amplitudes. We use a standard normal-mode solver, MINEOS \cite{mineos2011}, to calculate the frequency and amplitude perturbations under small perturbations of the 1D model. We first introduce the baseline lunar model chosen for the practical calculation in this work, and then show the results of linearity checks and comparisons with MINEOS.

We choose a simple but representative 5-layer spherically symmetric model similar to the one we chose in our previous work \cite{PhysRevD.109.064092,2025PhRvD.111f3014Z}, including: inner core (``IC'', $0-352$ km), outer core (``OC'', $353-480$ km), mantle (``M'', $481-1709$ km), crust (``C'', $1709.1-1725$ km), and surface (``S'', $1725.1-1737.1$ km). Within each layer, all physical parameters ($v_p,v_s,~\rho,~Q_{p},Q_{s}$ \footnote{Note: $\kappa=\rho\left(v_{p}^{2}-\frac{4}{3} v_{s}^{2}\right)$, $\mu = \rho v_{s}^{2}$}) remain the same. Between two adjacent layers the parameters are linearly interpolated. Original parameters of the baseline model are recorded in Table \ref{tab:model}, while the interpolated model can be found in ``InterpolatedModel'' in \cite{webs}. Radial refinement and interpolation of the initial model are intended solely to improve the accuracy and stability of the eigensolvers such as MINEOS. Once the refinement reaches a sufficient level, its finer details no longer affect the numerical results or the conclusions of this paper \footnote{Previous work \cite{PhysRevD.109.064092} has validated the appropriateness of this refinement-and-interpolation procedure.}.

\begin{table}[h]
\caption{\label{tab:model}Baseline lunar model. IC = inner core, OC = outer core, M = mantle, C = crust, S = surface}
\begin{ruledtabular}
\begin{tabular}{lcccccr}
& $r$(km) & $\rho$(g/cm$^3$) & $v_p$(km/s) & $v_s$(km/s) & $Q_p$ & $Q_s$ \\
\hline
IC & 0-352 & 7.76 & 4.3 & 2.3 &1000 &1000 \\
OC & 353-480 & 3.4 & 8.2 & 3.2 &675 &300 \\
M & 481-1709 & 3.4 & 7.9 & 4.5 &3400 &1500 \\
C & 1709.1-1725 & 2.762 & 5.5 & 3.3 &6750 &6750 \\
S & 1725.1-1737.1 & 2.762 & 3.2 & 1.8 &6750 &6750 \\
\end{tabular}
\end{ruledtabular}
\end{table}

We test the perturbation results from our theoretical formalism against those obtained with MINEOS. Comparisons are done mainly for amplitude observables ($O_2$ and $O_3$) rather than eigenfrequency observables ($O_1$), because the latter have been widely investigated in past geophysics research. The result of a representative case, a shear-wave speed perturbation of 0.2\% in the inner core, is plotted in Figure~\ref{fig:ICvs}. Other results are shown in the Supplemental Material \cite{supplemental}. From these comparisons, we find that the theoretical formulation agrees remarkably well with the numerical results (MINEOS) across several orders of magnitude.

\begin{figure}[htbp] 
    \centering
    
    \begin{subfigure}{}
        \includegraphics[width=\linewidth]{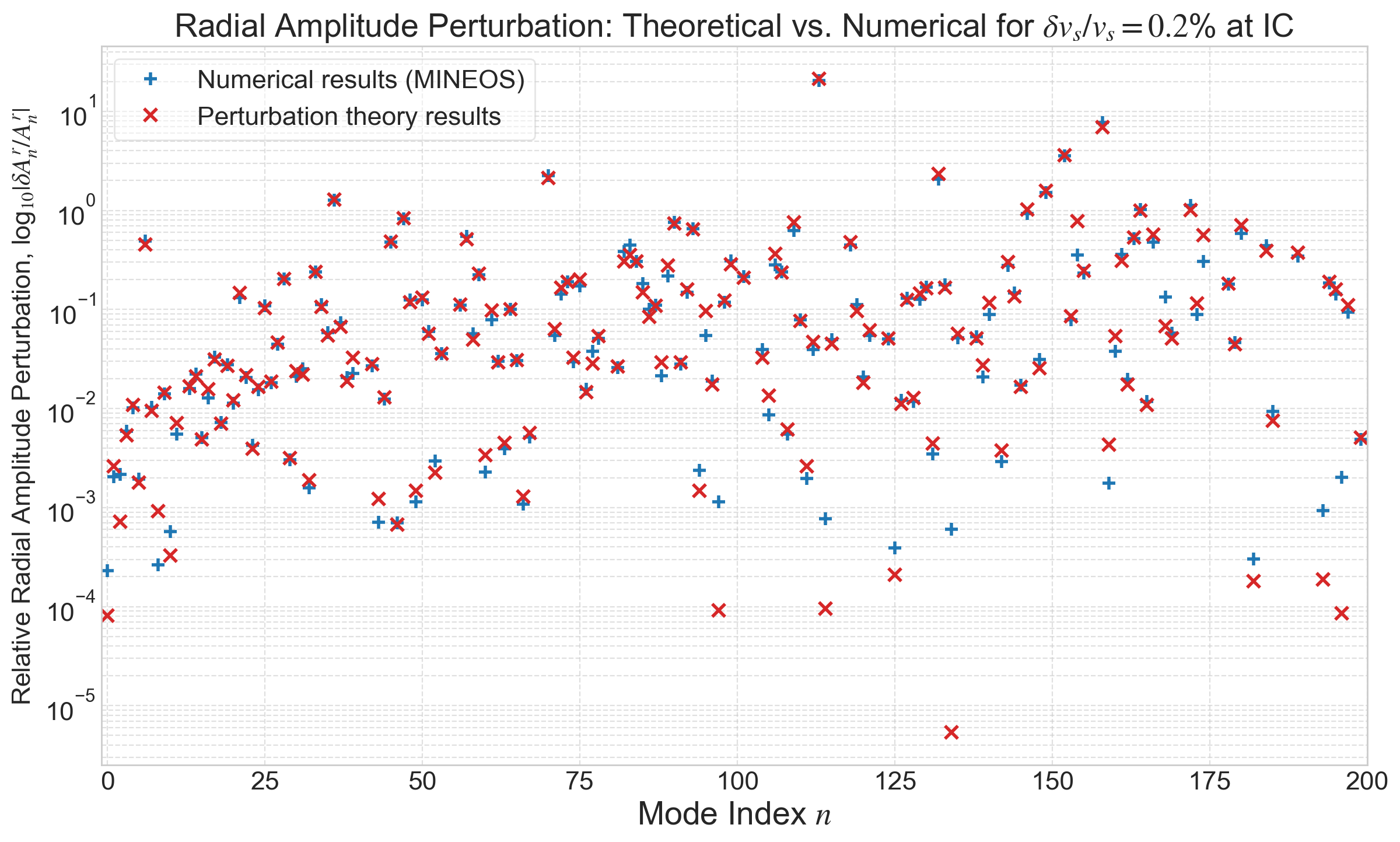}
    \end{subfigure}
    \begin{subfigure}{}
        \includegraphics[width=\linewidth]{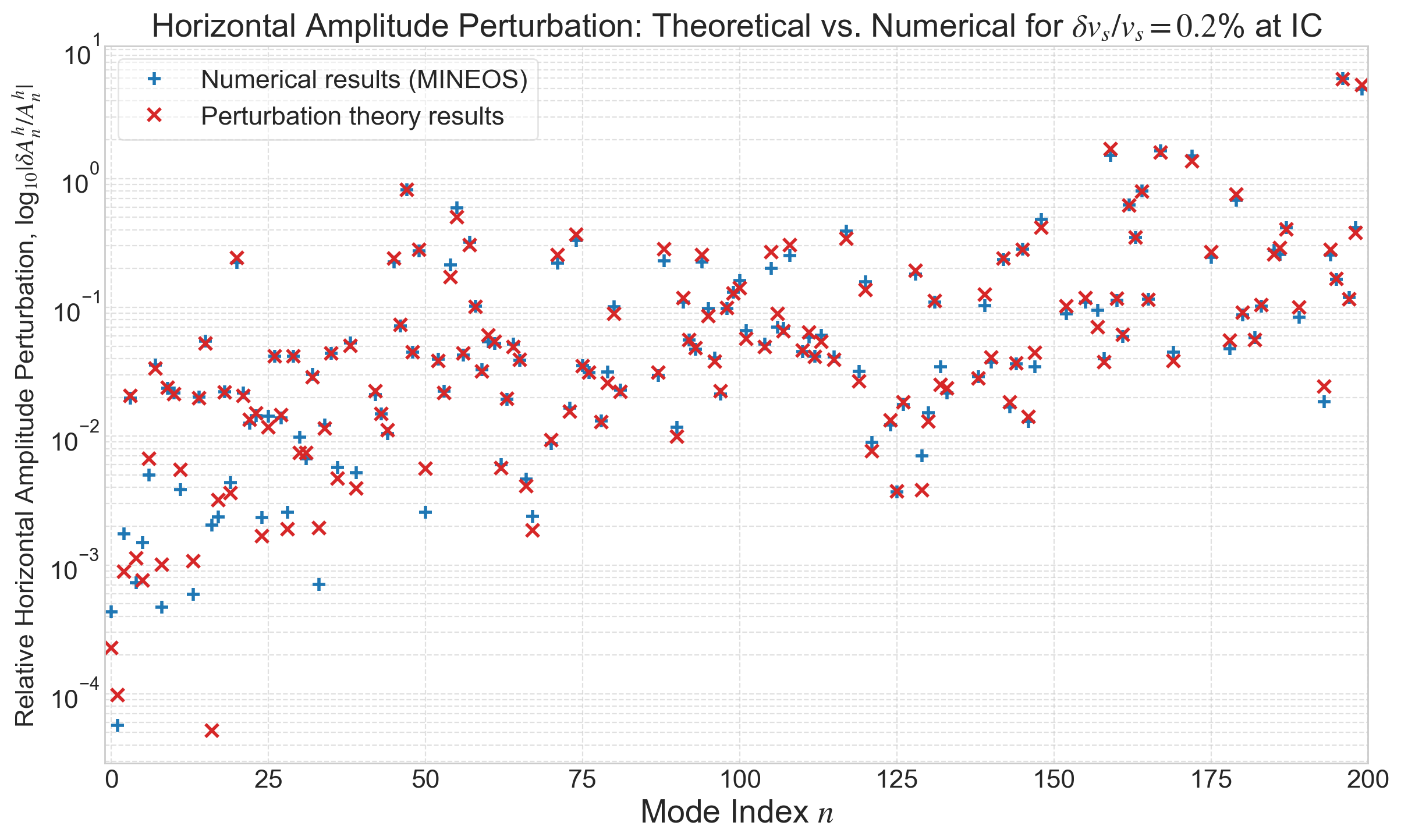}
    \end{subfigure}
    
    \caption{MINEOS verifications for amplitude perturbation, $\delta v_s / v_s = 0.2\%$ in the inner core. Only modes within the linear range (25\% tolerance of deviation) are included.}
    \label{fig:ICvs}
    
\end{figure}

\textit{Numerical example for GW tomography.}---Using our lunar baseline model, we perform a Fisher analysis for a comprehensive set of parameters and observables.  We consider three parameter types ($\kappa,\,\mu,\,\rho$) across different piecewise radial layers (based on the preliminary model but with increased radial resolution around $100~$km for most layers, as shown in Table~\ref{tab:lunar_layers} \footnote{Since we have a sufficient number of observables, far exceeding the number of model parameters, the specific layer thickness does not have significant influence on the results.}), and analyze a large suite of observables: eigenfrequencies, radial amplitudes and horizontal amplitudes for $0\le n\le150$ ($f_{150}\simeq 103.8~$mHz) under various combinations.  The perturbation summations ($\sum_{n'\neq n}$) include all modes with $n'\le200$ \footnote{Contributions of higher modes decay rapidly as $\omega_n^{-2}$.}.  We fix $\varepsilon_\omega=10^{-4}$ and parameterize the uncertainty of the amplitude as $\varepsilon_A=\alpha\,\varepsilon_\omega$. We require $\alpha \gg 1$ in this work.

\begin{figure*}[htbp] 
    \centering
    \includegraphics[width=0.95\linewidth]{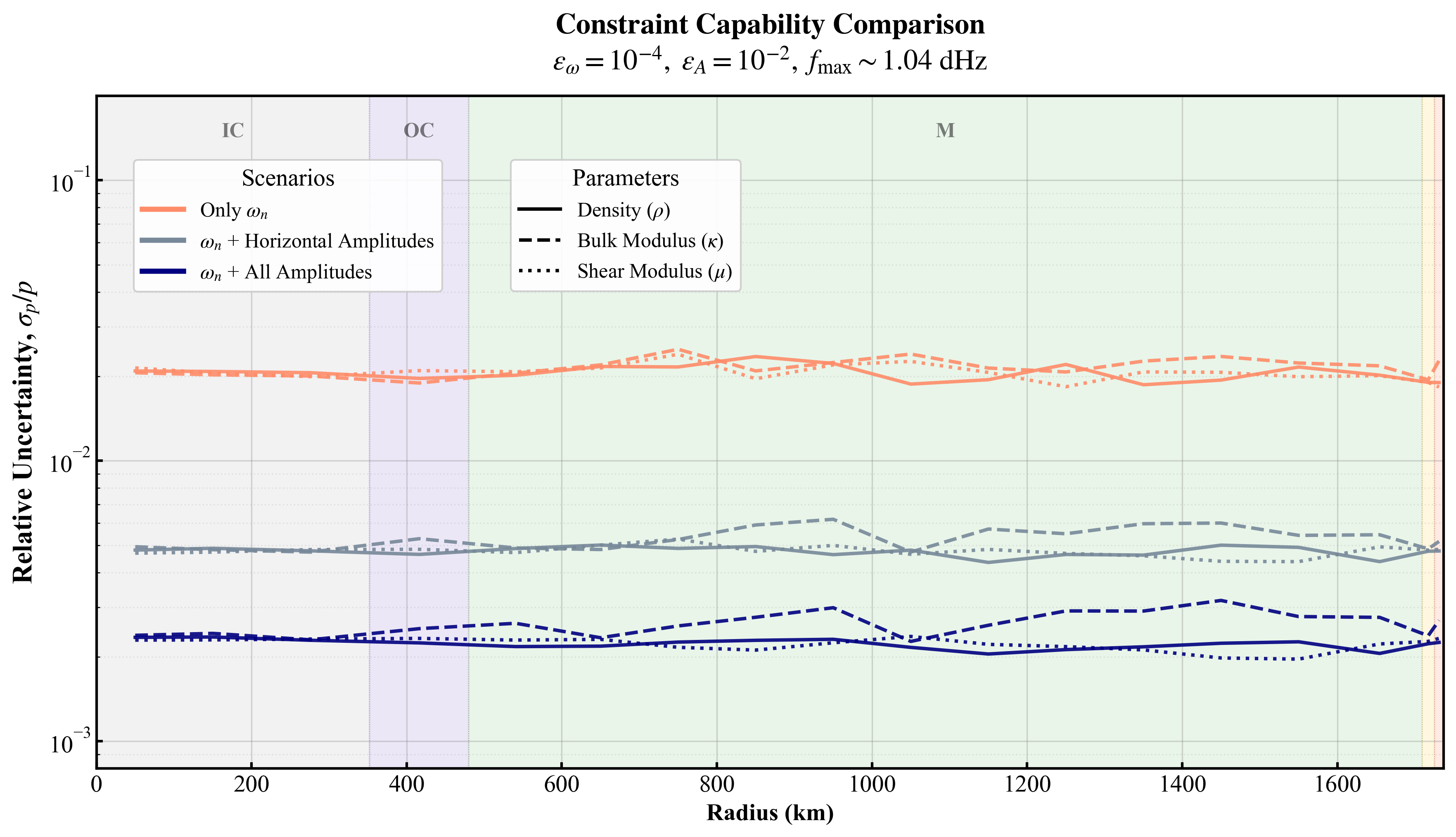}
    \caption{Relative errors for $\rho,\kappa,\mu$ with different combinations of observables. All modes with $n \le 150$ are considered as observables, while the perturbation summations are done for modes $\le 200$.}
    \label{fig:sigma}
\end{figure*}

Figure~\ref{fig:sigma} shows results for $\alpha=100$ and three observable combinations: eigenfrequency only, frequency plus horizontal amplitude (the LGWA case), and frequency plus all amplitudes. We see that the relative uncertainties for the three observational scenarios are weakly dependent on radius, which is partly because we only have tens of layers (Table~\ref{tab:lunar_layers}) while employing hundreds of observables. The results indicate that even when the amplitude measurements are two orders of magnitude less precise than the frequency measurements (i.e., \ $\alpha=100$), including the modal amplitudes in the data set improves the model constraints by nearly an order of magnitude for all three parameters $\rho,\kappa,\mu$. This improvement reflects a much better sensitivity of the modal amplitudes, rather than eigenfrequencies, to the model parameters, which can be mainly attributed to the $\frac{1}{\omega_k^2 - \omega_{k'}^2}$ factor in Eq.~(\ref{eq:delta_s_proj}).

\textit{Conclusion and discussions.}---In conclusion, we have presented a compact but operational framework for ``GW tomography'' of the Moon within the 1D (spherically symmetric) and linear approximation.  The framework explicitly links GW-driven modal amplitudes and eigenfrequency shifts to small perturbations of interior structure through Woodhouse kernels and first-order perturbation theory. Taken together, the analytical formulation and the Fisher pipeline provide a practical recipe to translate instrument and observational assumptions into quantitative statements about which physical parameters (bulk modulus, shear modulus, density, and interface locations) are in principle recoverable from GW-driven seismic data.

We summarize our major conclusions and points for follow-up as follows:
\begin{itemize}
  \item \textbf{Complementarity of frequency and amplitude measurements.} Eigenfrequency shifts and GW-induced modal amplitudes respond to different combinations of model perturbations. Because these two types of observables are sensitive to different aspects of the eigenfunctions, jointly inverting frequency and amplitude measurements can substantially increase the effective degrees of freedom and reduce parameter covariances compared with using either observable alone.  The quantitative benefit depends on the assumed observational precisions $\varepsilon_\omega$ and $\varepsilon_A$ ($\varepsilon_{\xi}$), the set of modes available for analysis, and the imposed priors. Practical diagnostics should be used to quantify these gains for any particular mission design.
  
  \item \textbf{Utilizing GWs to study solar-system phenomena.} An important conceptual point, perhaps the most novel implication of this work, is that GW-driven modal amplitudes, when the incident GW amplitude is calibrated by a detector network, become independent observables of the lunar interior structure. The tighter constraints on elastic parameters as a function of depth will greatly enhance our understanding of the lunar interior and might lead to breakthroughs in our understanding of the formation history of the Earth-Moon system \cite{NAP27209,2025JCAP...01..108A}. 
  
  \item \textbf{Near degeneracy.} Eigenmodes with similar eigenfrequencies naturally exhibit near-degeneracy effects, leading to greater errors or even failure of first-order perturbation theory. This problem can be overcome by applying a generalized version of the perturbation formalism: specifically, by using a larger subspace and solving the eigenvalue equation for a group of nearby eigenmodes \cite{1999tgs..book.....D,2001GeoJI.146..833D}.

  \item \textbf{Validity of the statistical and physical framework.} In this proof-of-principle study, we employ a 1D linearized approach and the Fisher information matrix (FIM) to project parameter uncertainties. While the FIM can be a poor error predictor in regimes with low signal-to-noise ratios or severe parameter degeneracies \cite{2008PhRvD..77d2001V}, its application is mathematically well-justified within our current framework. First, we adopt a relatively high expected SNR of around 100, which is achievable in future multi-band GW observations. This ensures that the statistical uncertainties remain small and the linearized-signal approximation remains fundamentally valid. Second, our simplified piecewise 1D structural model maintains a parameter dimensionality that is vastly outnumbered by the available observables (amplitudes at hundreds of resonance peaks across a broad frequency band). This heavily over-constrained setup naturally mitigates strong parameter correlations and yields a well-conditioned Fisher matrix, avoiding numerical instability and the severe amplification of errors during inversion.

  \item \textbf{Limitations and extensions.} Nevertheless, future realistic inversions must address the physical and statistical limitations of the current framework. The 1D, linearized approach neglects lateral heterogeneity and higher-order effects that may be relevant for large perturbations or for modes strongly affected by asphericity and topography. Extensions to 3D tomography will require aspherical sensitivity kernels (which are already included in Woodhouse's formalism) and a substantial increase in parameter dimensionality. Furthermore, in practical observational scenarios, the number of measurable modes may be lower, and the actual SNR might not reach the ideal thresholds assumed here, potentially invalidating the linear approximation. To address weakly nonlinear effects and increased parameter dimensionality, analytical Jacobians must be substituted with iterative Gauss–Newton or full nonlinear Bayesian inference schemes that systematically incorporate prior errors of the baseline models. Practical inversions will also need to address uncertainties in damping (mode $Q$ values) and explore joint inversions that combine GW-driven observables with independent lunar datasets (traditional seismic catalogs, tidal Love numbers, and geodetic constraints). Addressing these points will clarify the potential and limitations of GW tomography of the Moon under more realistic models and enable concrete recommendations for future lunar geophysical exploration.
\end{itemize}

\section*{Acknowledgments}
We thank Jeroen Tromp, Xian Chen, Luqian Jiang, Li Zhao, Yanbin Wang, Lei Zhang and Jinhai Zhang for many helpful discussions. This work is supported by the National Key Research and Development Program of China (Grant No.~2024YFC2207300) and the Italian Space Agency (ASI) under Grant No.~2025-29-HH.0. Han Yan acknowledges support from the China Scholarship Council (No.~202506010256). This work was carried out in parts under the ACME project funded by the European Union's Horizon Europe Research and Innovation Programme under Grant Agreement No. 101131928.

\bibliography{main}

\clearpage
\appendix
\section*{Supplemental Material}
\setcounter{equation}{0}
\setcounter{figure}{0}
\setcounter{table}{0}
\setcounter{section}{0}
\renewcommand{\thefigure}{A\arabic{figure}}
\renewcommand{\theequation}{A\arabic{equation}}
\renewcommand{\thetable}{A\arabic{table}}

\section{Theoretical details}
\subsection{Eigenfunctions and Normalization}

The vector eigenfunction is defined as
\begin{eqnarray}\label{eq:vecSpher}
        \mathbf{s}_{nlm}^{\sigma}= &&\delta_{\sigma S}[U_{n l}(r) \mathbf{P}_{l m}+V_{n l}(r) \mathbf{B}_{l m}] \nonumber \\
        &&+\delta_{\sigma T}[W_{n l}(r) \mathbf{C}_{l m}] ~,
\end{eqnarray}
where the vector spherical harmonics ($\mathbf{P}_{l m},~\mathbf{B}_{l m}$ and $\mathbf{C}_{l m}$) follow the definition in \textit{Theoretical Global Seismology} \cite{1999tgs..book.....D}. For a spheroidal mode, Eq.~(\ref{eq:vecSpher}) can be simplified to
\begin{eqnarray}
    \mathbf{s}_{nlm}^{S} =&& U_{n l}(r) \mathbf{P}_{l m}+V_{n l}(r) \mathbf{B}_{l m} \nonumber \\
    =&&  U_{n l}(r) \hat{\mathbf{r}} \mathcal{Y}_{lm}  +\nonumber \\
    &&   \frac{V_{n l}(r)}{\sqrt{l(l+1)}} [\hat{\boldsymbol{\theta}} \partial_{\theta}+\hat{\boldsymbol{\phi}}(\sin \theta)^{-1} \partial_{\phi}] \mathcal{Y}_{l m} ~.
\end{eqnarray}
Therefore, $U_{n l}$ represents the radial response, and $V_{n l}$ represents the horizontal response.
The normalization condition is given by
\begin{align}
    \int_{0}^{R} \rho (U_{k}U_{k'}+V_{k}V_{k'})r^2 \mathrm{d}r &= \delta_{kk'} ~,\nonumber \\
     \int_{0}^{R} \rho W_{k}W_{k'}r^2 \mathrm{d}r &= \delta_{kk'} ~.
\end{align}

\subsection{Overlap Integral and Kernels}

The overlap integral $I_n$ governing the GW coupling strength in Eq.~(\ref{eq:mode_amp}) can be expressed explicitly as \cite{2019PhRvD.100d4048M}:
\begin{equation}\label{eq:I_n}
  I_n \equiv \int_0^R \partial_{r}\mu(r)\,\left[U_n(r)+\frac{3}{\sqrt{6}}V_n(r)\right] r^2 \mathrm{d}r~,
\end{equation}
under the assumption of a Dyson-type force density \cite{1969ApJ...156..529D,1983NCimC...6...49B,PhysRevD.109.064092}.

The perturbation matrix elements in Eq.~(\ref{eq:delta_s_proj}) are computed via radial integrals over Woodhouse kernels $V_p^{n'n}(r),~T_p^{n'n}(r)$. For spherically symmetric perturbations and a parameter $p \in \{\rho, \kappa, \mu\}$:
\begin{align}\label{eq:WoodhouseKernel}
  \langle\mathbf{s}_{k m}|\delta V [\delta p] - \omega_k^2 \delta T [\delta p]|\mathbf{s}_{k' m'}\rangle & = \delta_{ll'}\delta_{mm'}\sqrt{\frac{2l+1}{4\pi}} \nonumber \\
  & \times\int_0^R r^2\,\delta p(r)\,K_p^{n'n}(r)\,\mathrm{d}r ~,
\end{align}
where $K_p^{n'n} = V_p^{n'n}-\omega_k^2 T_p^{n'n}$.
Explicit expressions for these kernels follow the definition in \cite{1999tgs..book.....D}. For an interface perturbation at boundary $d_i$, $r^2 \delta p \mathrm{d}r$ should be replaced by $d_i^2 \delta d_i$.

\subsection{Partial Derivatives (Jacobians)}
\label{sec:deriv}

Although in a high-resolution model $p$ should be regarded as a function of radius [e.g., $\rho(r)$], in a preliminary and realistic calculation, $p$ can be treated as a piecewise constant function, which is also common in geophysics research. In this work, we choose a piecewise baseline model (introduced in Table \ref{tab:model}), and the parameter perturbations are also regarded as piecewise constant functions but with smaller radial segment lengths (see Table~\ref{tab:lunar_layers}).

For a localized perturbation $\delta p(r) = \delta p$ on $r\in[r_1,r_2]$, i.e., assuming the perturbations of these functions to be constant within a given radius range, the functional derivatives entering a linearized inversion are:

\emph{Eigenfrequency derivative:}
\begin{align}
  \frac{\partial (\delta\omega_n^2)}{\partial (\delta p)}
  =& \sqrt{\frac{2l+1}{4\pi}}\;
 \int_{r_1}^{r_2} \mathrm{d}r' r'^2\; \bigl[V_p^{nn}(r')-\omega_k^2 T_p^{nn}(r')\bigr]~. \label{eq:kernel_omega}
\end{align}

\emph{Eigenfunction derivative:}
The variation of the scalar functions is derived from Eq.~(\ref{eq:delta_s_proj}):
\begin{align}
    \frac{\partial (\delta U_n)}{\partial (\delta p)} (r)
  =& \sqrt{\frac{2l+1}{4\pi}}\;
 \sum_{n'\neq n} U_{n'}(r)  \nonumber \\
  &\times\int_{r_1}^{r_2} \mathrm{d}r' r'^2\;
    \frac{\bigl[V_p^{n'n}(r')-\omega_n^2T_p^{n'n}(r')\bigr]}{\omega_n^2-\omega_{n'}^2} ~. \label{eq:kernel_s}
\end{align}
The expression for $\partial (\delta V_n) / \partial (\delta p)$ is strictly analogous.

\emph{Overlap integral derivative:}
The derivative of $I_n$ is:
\begin{align}\label{eq:det_In}
    \frac{\partial (\delta I_n)}{\partial (\delta p)} 
  =& \sqrt{\frac{2l+1}{4\pi}}\;
 \sum_{n'\neq n} I_{n'}  \nonumber \\
  &\times\int_{r_1}^{r_2} \mathrm{d}r' r'^2\;
 \frac{\bigl[V_p^{n'n}(r')-\omega_n^2T_p^{n'n}(r')\bigr]}{\omega_n^2-\omega_{n'}^2} \nonumber \\
    &+\Bigg[ \int_{r_1}^{r_2} \partial_{r'}\delta \mu(r') \left(U_n(r)+\frac{3}{\sqrt{6}}V_n\right) r^2 \mathrm{d}r \Bigg]_{p=\mu}~.
\end{align}
The variations in surface amplitude observables $O_{2,n}$ and $O_{3,n}$ depend directly on the variations of $U_n(R), V_n(R)$ and $I_n$. As a result, the above partial derivatives naturally give the elements of the Jacobian matrix $G$. For example, we have
\begin{eqnarray}
    \frac{1}{\sigma_{2,n}} \frac{\partial O_{2,n}}{\partial p[r_1,r_2]} \simeq \frac{1}{\varepsilon_A} &&  \bigg[ \frac{1}{U_n (R)}\frac{\partial (\delta U_n)}{\partial (\delta p)} (R) \nonumber \\
    &&+ \frac{1}{I_n} \frac{\partial (\delta I_n)}{\partial (\delta p)} \bigg]  ~.
\end{eqnarray}

\section{Linearity check of the perturbation}
\label{app:linearity} 

\begin{figure}[htbp] 
    \centering
    
    \begin{subfigure}{}
        \includegraphics[width=0.98\linewidth]{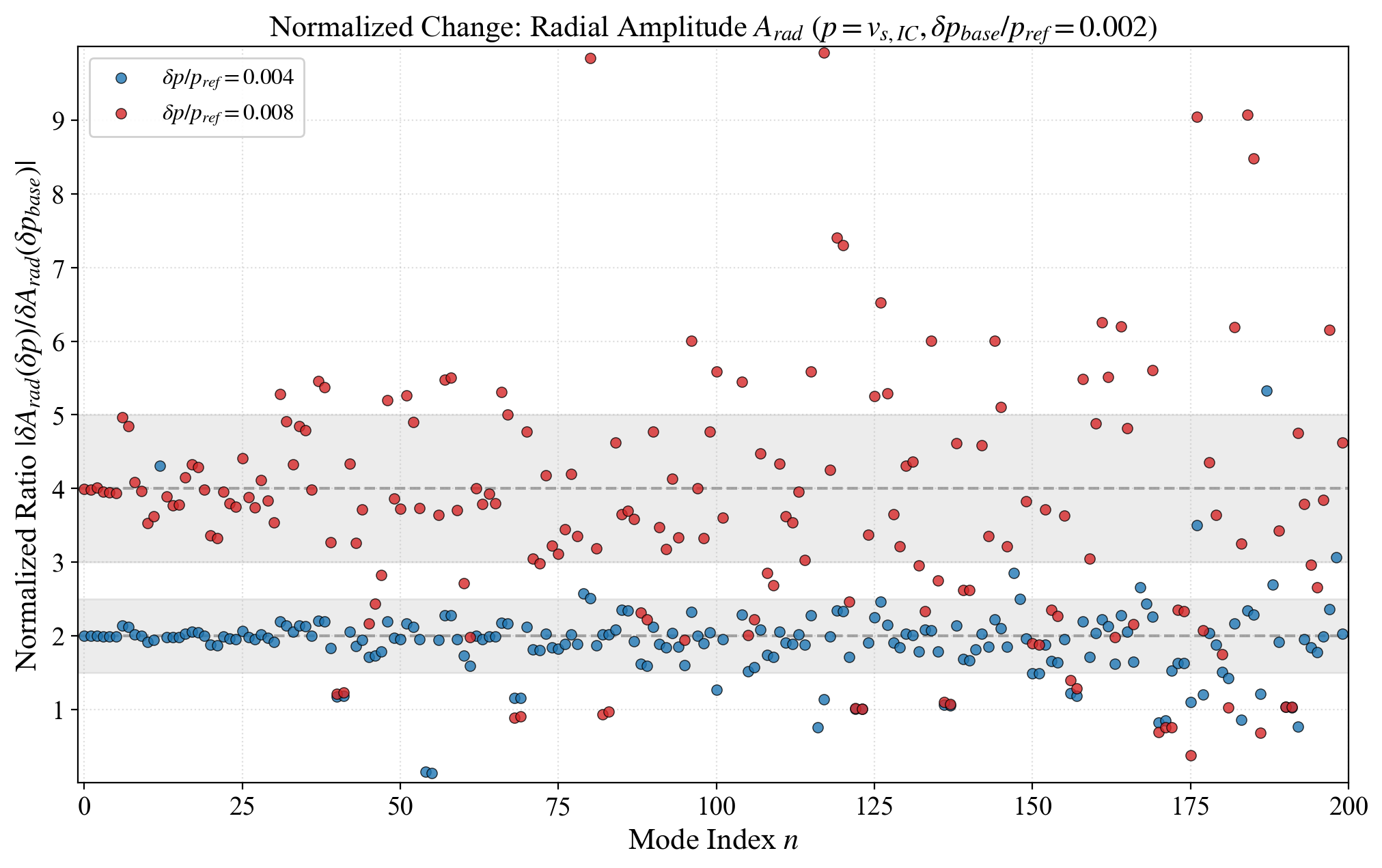}
    \end{subfigure}
    \begin{subfigure}{}
        \includegraphics[width=0.98\linewidth]{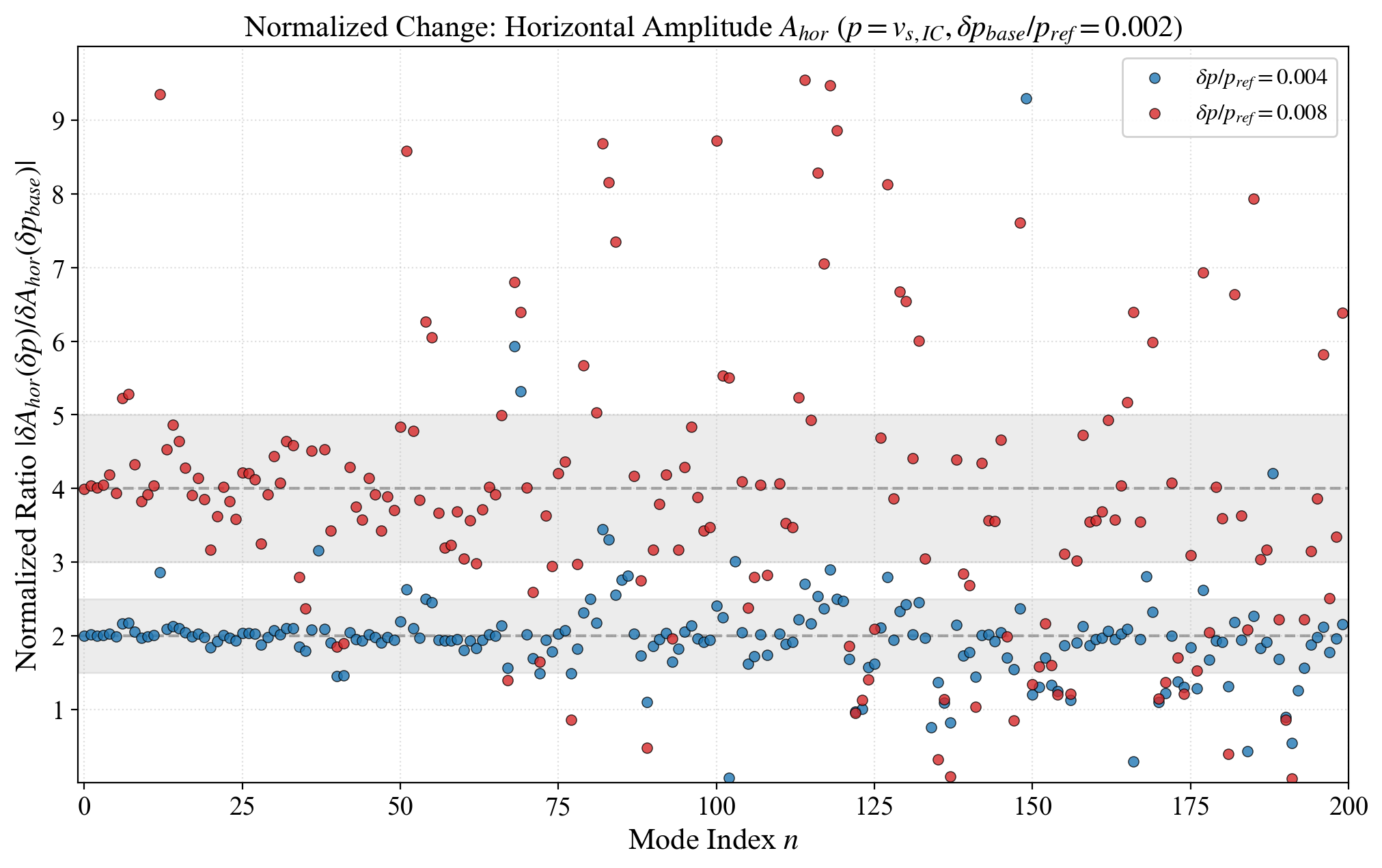}
    \end{subfigure}
    
    \caption{MINEOS examination for linearity range of $A(\delta v_s ) /A(\delta v_{s,base} )$ in the inner core.}
    \label{fig:ICvs-linear}
    
\end{figure}

\begin{figure}[htbp] 
    \centering
    
    \begin{subfigure}{}
        \includegraphics[width=0.98\linewidth]{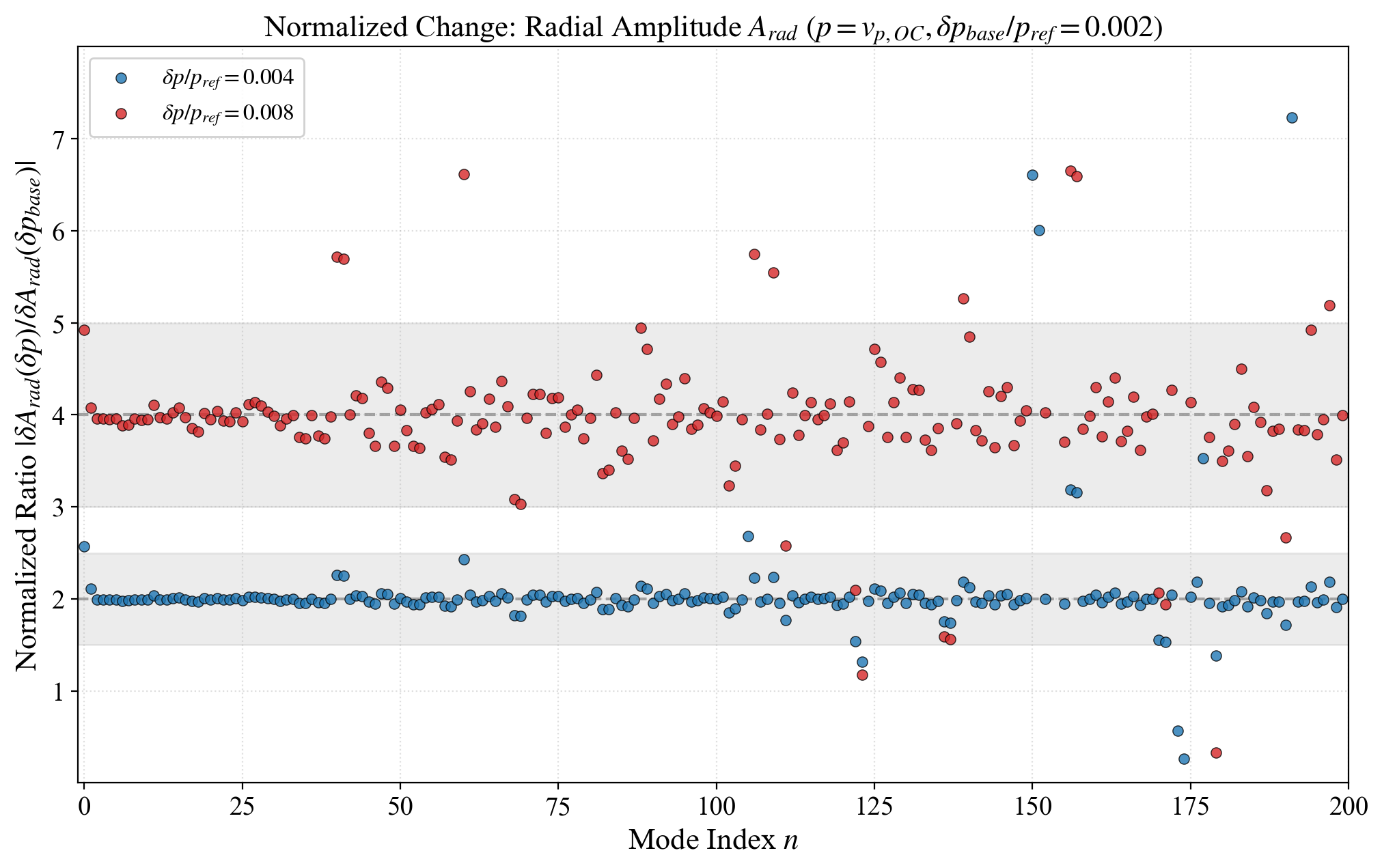}
    \end{subfigure}
    \begin{subfigure}{}
        \includegraphics[width=0.98\linewidth]{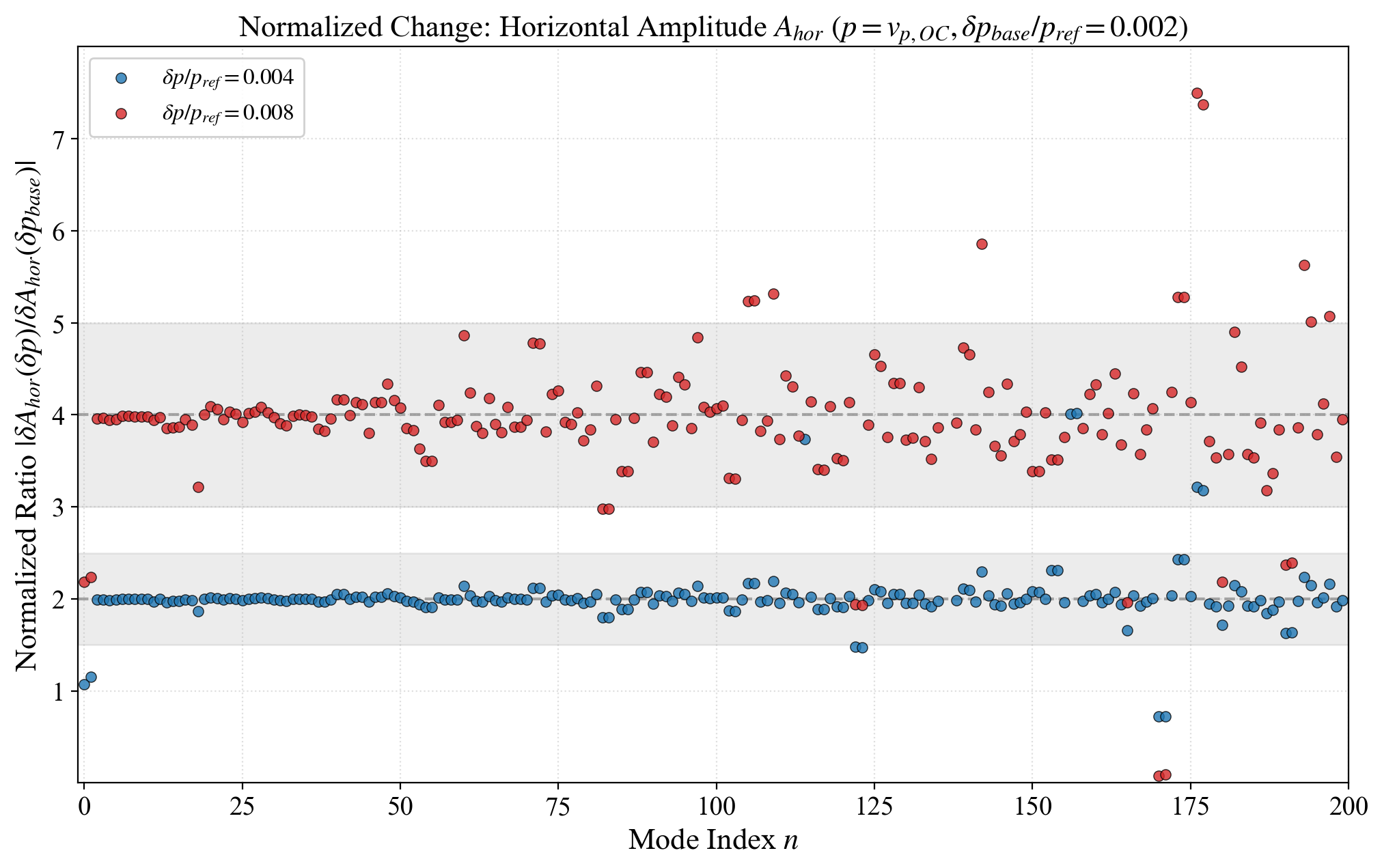}
    \end{subfigure}
    
    \caption{MINEOS examination for linearity range of $A(\delta v_p ) /A(\delta v_{p,base} )$ in the outer core.}
    \label{fig:OCvp-linear}
    
\end{figure}

\begin{figure}[htbp] 
    \centering
    
    \begin{subfigure}{}
        \includegraphics[width=0.98\linewidth]{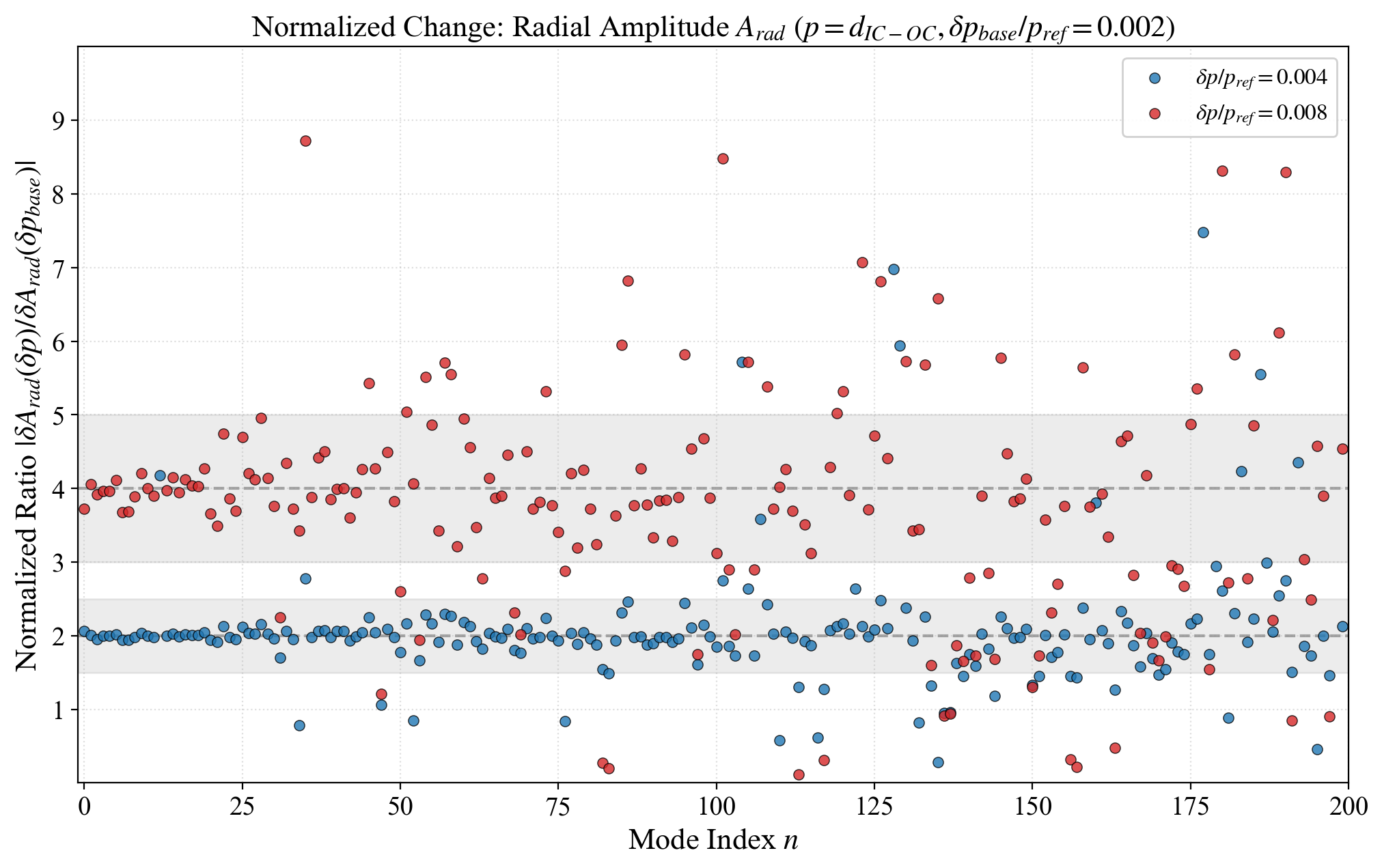}
    \end{subfigure}
    \begin{subfigure}{}
        \includegraphics[width=0.98\linewidth]{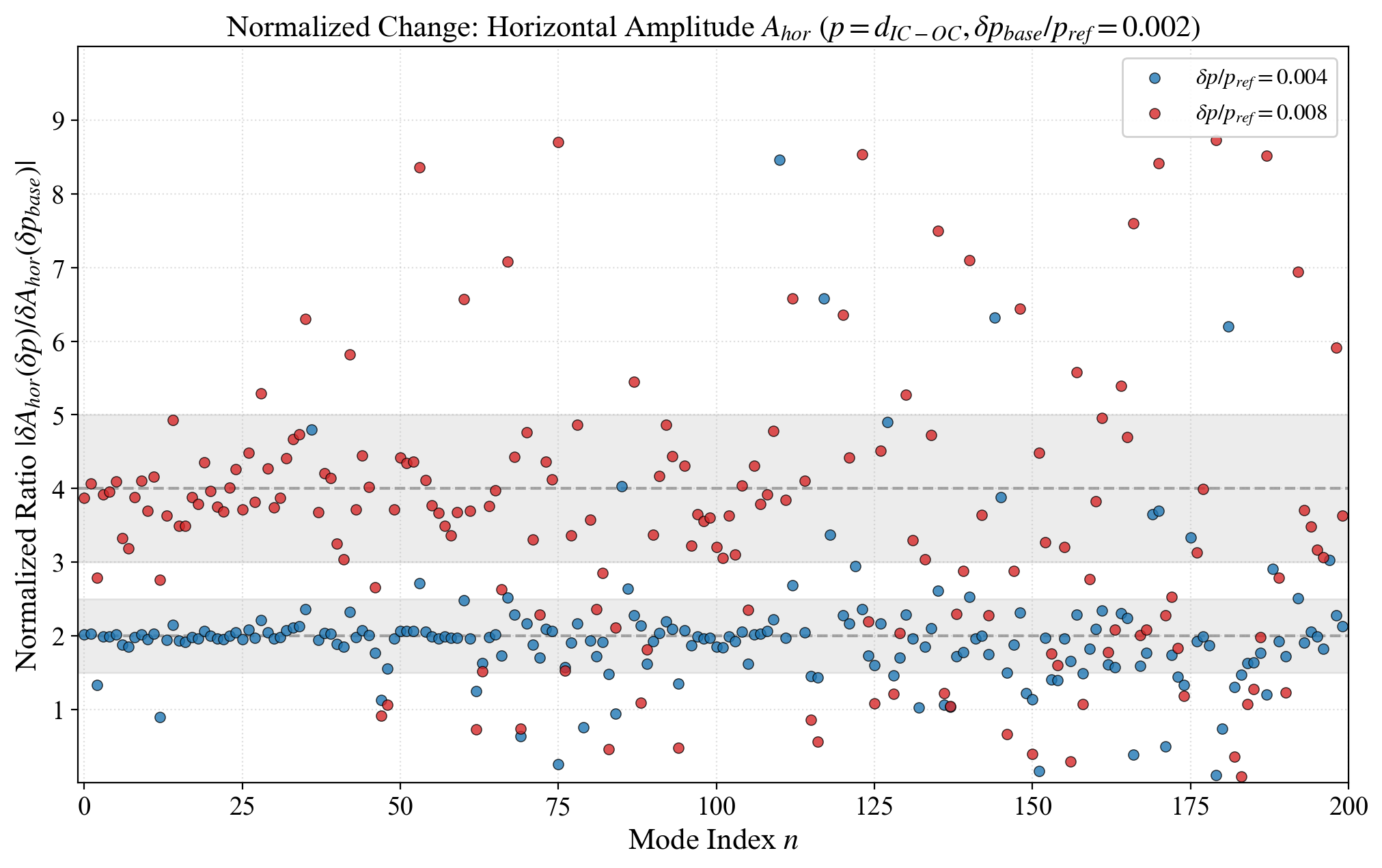}
    \end{subfigure}
    
    \caption{MINEOS examination for linearity range of $A(\delta d_i ) /A(\delta d_{i,base} )$ at the boundary between inner core and outer core.}
    \label{fig:ICOC-linear}
    
\end{figure}

In this section, we examine the linearity of the amplitude response to the parameter perturbation by using the MINEOS results for 0.2\%, 0.4\%, and 0.8\% parameter perturbations. We perform MINEOS calculations for inner-core shear-wave velocity perturbation (Fig.~\ref{fig:ICvs-linear}), outer-core compression-wave velocity perturbation (Fig.~\ref{fig:OCvp-linear}), and the interface radius perturbation between the inner core and outer core (Fig.~\ref{fig:ICOC-linear}). All modes with $n \le 200$ are included. We plot the ratio of the amplitude variation relative to the ``base perturbation'' (0.2\%) in the plots, i.e.,
\begin{equation}
    \frac{\delta A (\delta p)}{\delta A (\delta p_{base})},~\frac{\delta p_{base}}{p_{ref}}=0.2\%~, \nonumber
\end{equation}
(where ``ref'' denotes the unperturbed reference model,) for a clear examination of the linearity. Horizontal gray dashed lines represent the theoretical values (i.e., $\frac{0.4\%}{0.2\%}=2$ and $\frac{0.8\%}{0.2\%}=4$) of this ratio. The gray stripes on both sides of the dashed line indicate the 25\% range of deviation, for clarity.

Generally, we find that most of the modes (over 75\%) are within the linearity range of 25\%.

In general, within the first 200 modes, most (around 70\%) show acceptable linearity. Deviation from linearity arises naturally from near-degeneracy effects, which can in principle be handled by a minor expansion of our formalism as discussed in the Letter. On the other hand, omitting near-degeneracy and simply including all modes generally underestimates the perturbation-theory results, indicating that we can directly deploy this simple version of perturbation calculation in this exploratory study.

\section{Examining the perturbation theory by MINEOS}
\label{app:examine} 

\begin{figure}[htbp] 
    \centering
    
    \begin{subfigure}{}
        \includegraphics[width=0.98\linewidth]{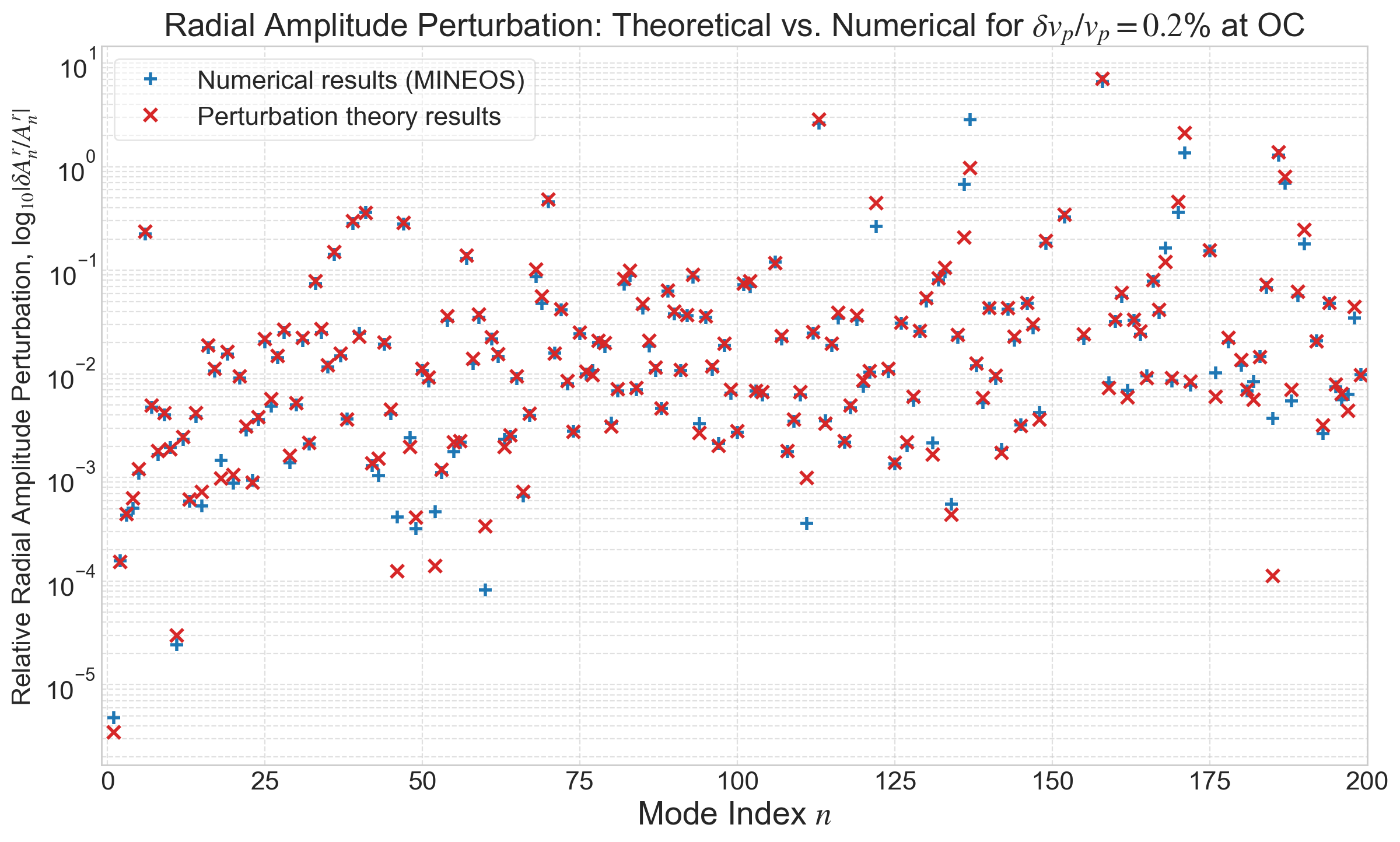}
    \end{subfigure}
    \begin{subfigure}{}
        \includegraphics[width=0.98\linewidth]{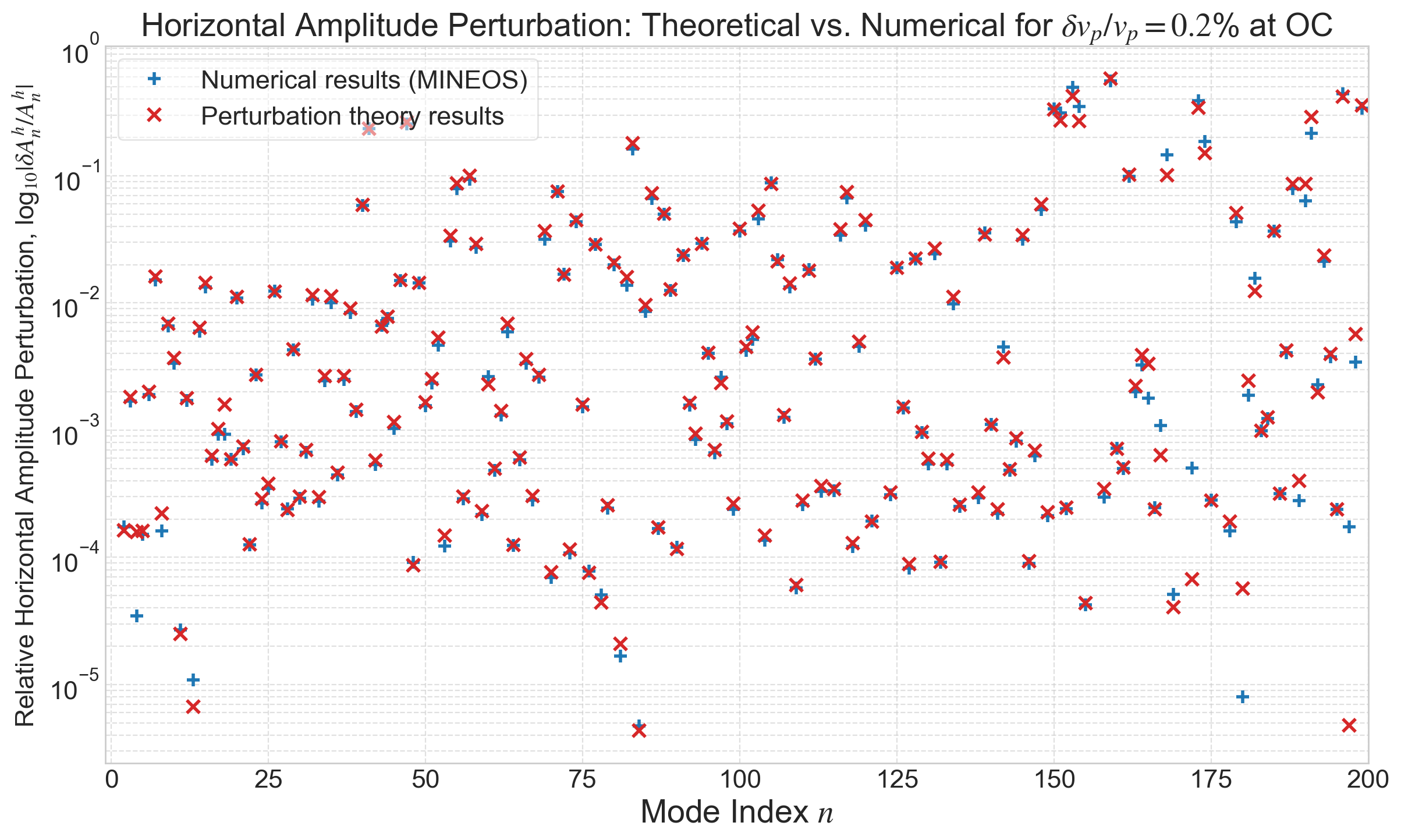}
    \end{subfigure}
    
    \caption{MINEOS verifications for amplitude perturbation, $\delta v_p / v_p = 0.2\%$ in the outer core. Only modes within the linear range are included.}
    \label{fig:OCvp}
    
\end{figure}

\begin{figure}[htbp] 
    \centering
    
    \begin{subfigure}{}
        \includegraphics[width=0.98\linewidth]{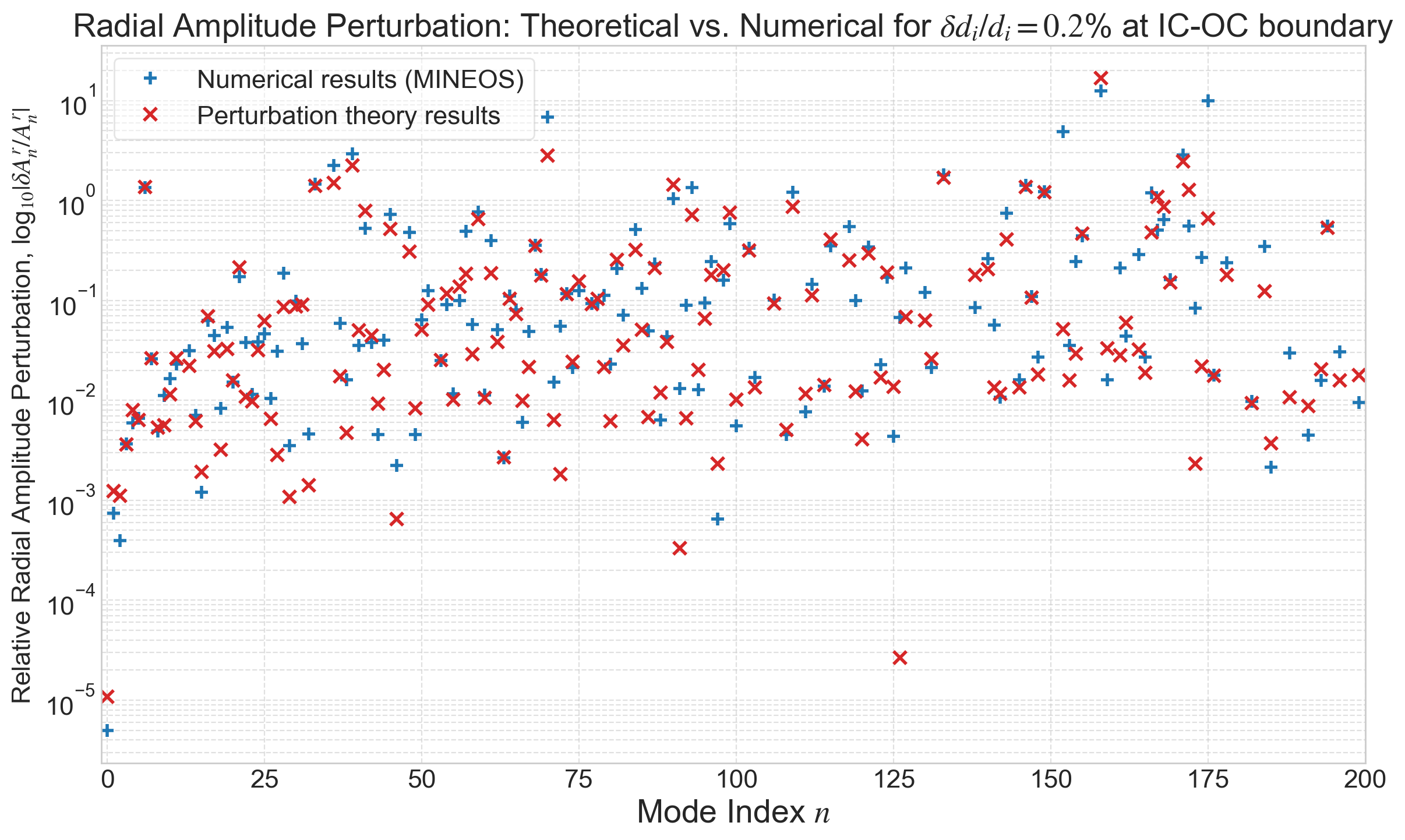}
    \end{subfigure}
    \begin{subfigure}{}
        \includegraphics[width=0.98\linewidth]{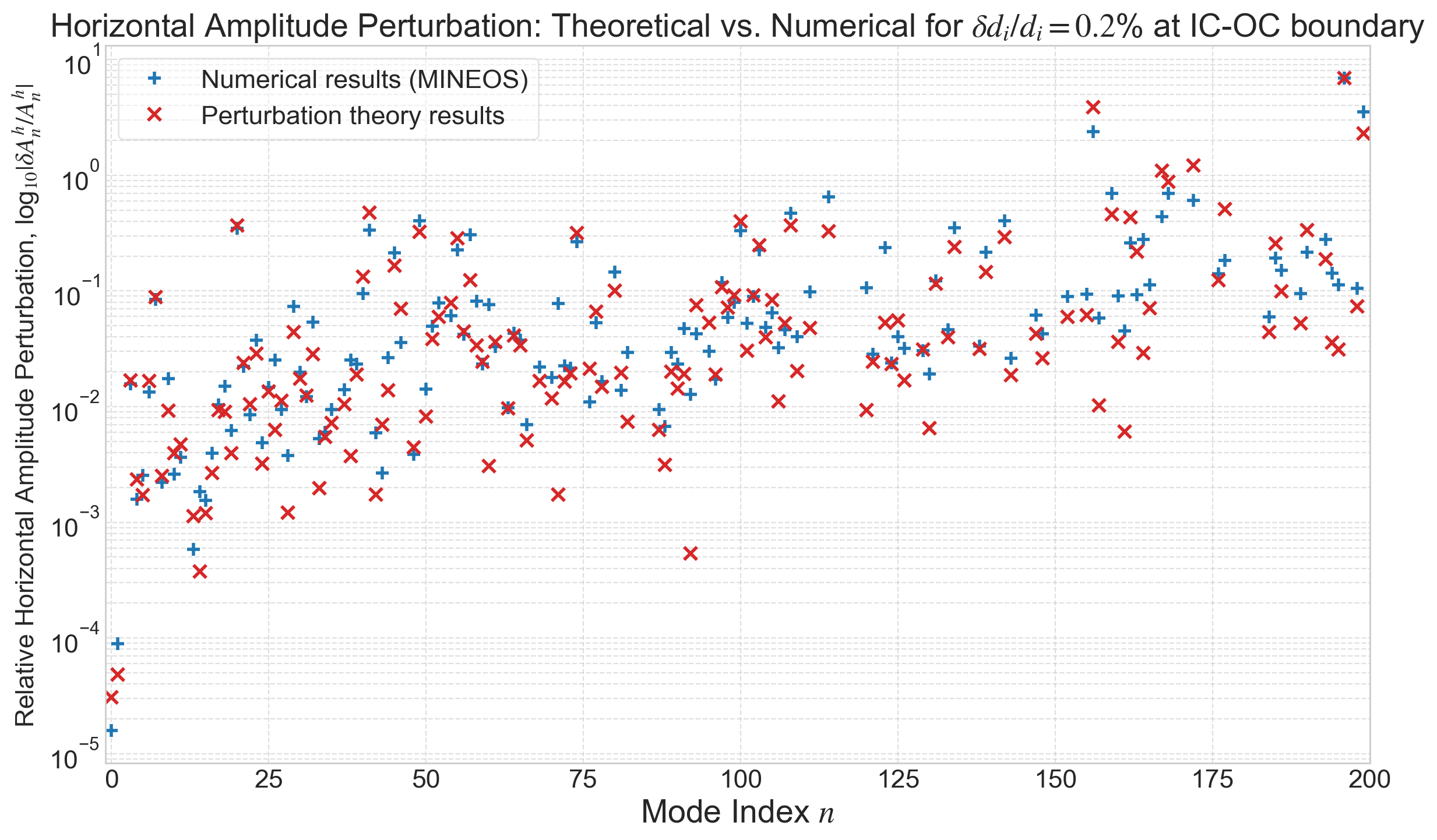}
    \end{subfigure}
    
    \caption{MINEOS verifications for amplitude perturbation, $\delta d_i / d_i = 0.2\%$ at IC-OC boundary. Only modes within the linear range are included.}
    \label{fig:IC-OC}
    
\end{figure}

We also compare the perturbation-theory results and the MINEOS results for 0.2\% compression-wave velocity perturbation in the outer core (Fig.~\ref{fig:OCvp}), and the interface radius perturbation between the inner core and outer core (Fig.~\ref{fig:IC-OC}). Modes within the linear range (25\% threshold) are included.

For shear-wave-velocity perturbations (Fig.~\ref{fig:ICvs}), minor discrepancies appear mainly for modes whose amplitudes are extremely small, because the term $\partial (\delta \mu)$ in Eq.~(\ref{eq:det_In}) has been omitted in this comparison. Besides, we also find larger discrepancies for interface perturbation, which is mainly due to the larger errors and instability of the density term's contribution in the coupling matrix. In any case, the discrepancies mentioned above will not change the qualitative conclusions of this work.

\clearpage
\section{Radial layers for parameter estimation}

Table~\ref{tab:lunar_layers} describes the locations of the piecewise radial layers for parameter estimation. The boundaries are based on the preliminary model in Table~\ref{tab:model} but with an increased radial resolution of about 100 km.

\begin{table}[htbp]
\caption{Radial layers for parameter estimation.}
\label{tab:lunar_layers}
\begin{ruledtabular}
\begin{tabular}{ccc}
Layer number & $r_1$ (km) & $r_2$ (km) \\
\colrule
1  & 0       & 100     \\
2  & 100     & 200     \\
3  & 200     & 352     \\
4  & 353     & 480     \\
5  & 481     & 600     \\
6  & 600     & 700     \\
7  & 700     & 800     \\
8  & 800     & 900     \\
9  & 900     & 1000    \\
10 & 1000    & 1100    \\
11 & 1100    & 1200    \\
12 & 1200    & 1300    \\
13 & 1300    & 1400    \\
14 & 1400    & 1500    \\
15 & 1500    & 1600    \\
16 & 1600    & 1709    \\
17 & 1709.1  & 1725    \\
18 & 1725.1  & 1737.1  \\
\end{tabular}
\end{ruledtabular}
\end{table}

\end{document}